# Providing Traceability for Neuroimaging Analyses

R. McClatchey, A. Branson, A. Anjum[1], P. Bloodsworth[2], I. Habib, K. Munir, J. Shamdasani, K. Soomro and the neuGRID Consortium[3]

Centre for Complex Cooperative Systems, CEMS Faculty, Univ. of the West of England, Coldharbour Lane, Frenchay, Bristol BS16 1QY, United Kingdom
Telephone: +44 117 328 3761, FAX: +44 117 344 3155
Emails: {Richard.McClatchey, Andrew.Branson, Ashiq.Anjum, Kamran.Munir, Jetendr.Shamdasani, Kamran.Soomro}@cern.ch; Irfan.Habib@uwe.ac.uk; Peter.Bloodsworth@seecs.edu.pk

## Abstract

**Introduction**

With the increasingly digital nature of biomedical data and as the complexity of analyses in medical research increases, the need for accurate information capture, traceability and accessibility has become crucial to medical researchers in the pursuance of their research goals. Grid- or Cloud-based technologies, often based on so-called Service Oriented Architectures (SOA), are increasingly being seen as viable solutions for managing distributed data and algorithms in the bio-medical domain. For neuroscientific analyses, especially those centred on complex image analysis, traceability of processes and datasets is essential but up to now this has not been captured in a manner that facilitates collaborative study.

**Purpose and Method**

Few examples exist, of deployed medical systems based on Grids that provide the traceability of research data needed to facilitate complex analyses and none have been evaluated in practice. Over the past decade, we have been working with mammographers, paediatricians and neuroscientists in three generations of projects to provide the data management and provenance services now required for 21st century medical research. This paper outlines the finding of a requirements study and a resulting system architecture for the production of services to support neuroscientific studies of biomarkers for Alzheimer's Disease.

**Results**

The paper proposes a software infrastructure and services that provide the foundation for such support. It introduces the use of the CRISTAL software to provide provenance management as one of a number of services delivered on a SOA, deployed to manage neuroimaging projects that have been studying biomarkers for Alzheimer's disease.

**Conclusions**

In the neuGRID and N4U projects a Provenance Service has been delivered that captures and reconstructs the workflow information needed to facilitate researchers in conducting neuroimaging analyses. The software enables neuroscientists to track the evolution of workflows and datasets. It also tracks the outcomes of various analyses and provides provenance traceability throughout the lifecycle of their studies. As the Provenance Service has been designed to be generic it can be applied across the medical domain as a reusable tool for supporting medical researchers thus providing communities of researchers for the first time with the necessary tools to conduct widely distributed collaborative programmes of medical analysis.

---

[1] Now at School of Computing and Mathematics, University of Derby, Derby, UK
[2] Now at School of Electrical Engineering and Computer Science, NUST, Islamabad, Pakistan
[3] Full set of clinicians and informaticians listed in Acknowledgements

# 1. Introduction

The last few years have seen massive increases in computing power and data storage capacity enabling new applications that can handle increasingly complex and large volumes of data. Advances in network speed have enabled applications to be distributed over the web, providing the potential for improved resource utilisation and on-demand sharing. Medical informatics is one domain where these technological advances can bring significant benefit both for scientific research and for day-to-day clinical provision. With the arrival of a deluge of digitised information that has resulted from advances in the medical domain, clinical research is faced with increasing problems of data management and provenance in data analysis.

Over the past two decades, Grid computing has emerged as a potential candidate for supporting large-scale experiments in bio-medical and other scientific domains. Grid computing can be defined as the "flexible, secure, coordinated resource sharing among dynamic collections of individuals, institutions and resources" [1]. The Grid and latterly the Cloud [2] have provided the infrastructures and platforms to address the research challenges in medical research (as examples see [3], [4], and [5]). Emphasis has now shifted from the development of such infrastructures, to the provision of services through which medical researchers can access data and algorithms to facilitate their programmes of research. As an example, consider computational neuroimaging research; it requires enormous computing resources and the availability of larger MRI datasets will further enhance the need for large-scale distributed processing and data management. Recently research effort has been focussed on providing large image repositories, e.g. the recent US Alzheimer Disease Neuroimaging Initiative (ADNI) [6]. Geographically distributed infrastructures for computational analyses have been established to enable the sharing of resources and intensive data analysis to advance knowledge of neurodegenerative diseases. Several projects, such as NeuroLOG [7] and Neurogrid [8], have been undertaken to provide Grid infrastructures that support neuroimaging applications.

The study of Alzheimer's disease was selected as the application domain for our work because it was an early adopter of imaging-based research techniques. The search for imaging biomarkers is a complex task and has led to the use of resource intensive image processing algorithms which measure physical brain features, such as the thickness of the cortex. Until recently such analyses could only be carried out locally on a high specification desktop or a local cluster. The growth in both the number of images becoming available via international studies such as ADNI and the increasing resolution of scans will make this local approach unsustainable in the near future. Many research groups cannot create large-scale computing infrastructures locally because of the cost, space and maintenance issues that are associated with such facilities.

The neuGRID project [9] ran from 2006 to 2011 and was an EC-funded infrastructure initiative, which facilitated the collection and archiving of large amounts of imaging data along with the provision of analysis services. It was followed by the N4U (neuGRID for You [10]) project (2011-2014) that provides user-facing services, including provenance services, to enable neuroimaging analyses to be performed using the data stored in the neuGRID infrastructure. The intended benefit of these projects is to enable the discovery of biomarkers for Alzheimer's disease that will improve diagnosis and help speed the development of innovative drugs. Computational power on its own however, is not sufficient to make the infrastructure useful for clinical researchers. In both of the neuGRID and N4U projects, the end-user research community has identified the vital need for data provenance. We have addressed this through the provision of a so-called Provenance Service and an Analysis Base, a description and evaluation of which are the main contributions of this article.

This paper aims to demonstrate to the medical informatics community a practical, computer supported implementation of medical analysis traceability and to indicate the benefits of provenance data management. Firstly we outline the infrastructures that support service-based neuroimaging analysis. We then investigate the need for provenance that is evident in the specification and execution of neuroimaging analysis workflows (or pipelines) and in the definition of data samples used in studies of Alzheimer's disease; this section also introduces the neuGRID Provenance Service. The following section describes the use of a system called CRISTAL, as the basis of the Provenance Service. The use of CRISTAL is evaluated as a practical use case in the penultimate section of the paper and we draw lessons on its use.

## 2. Methods and Background
### 2.1 Infrastructure and Architecture for Neuroscience Analyses

The design philosophy that underpins both neuGRID and N4U is based on the proven object-oriented computing principles of reuse, flexibility and expandability. A service-oriented approach has been followed in these projects to deliver against these design principles. In neuGRID the specific requirements of neuroscientists were collected and analysed and services to satisfy those requirements have been designed and implemented to be flexible in nature and reusable in application. As shown in Figure 1, a set of reusable and generic services has been developed, resulting in a system that facilitates neuroscience research but retains a large degree of underlying generality.

In computing, a service-oriented architecture (SOA [11]) represents an approach in which a loose coupling of functions between operating systems, programming languages and applications can be achieved; a SOA thereby separates functions into distinct services [12] for usage. The MammoGrid project was the first healthcare example of a Grid-based service-oriented architecture [13]. Its design philosophy was to provide a set of reusable, generic services that were independent both from the front-end clinician-facing software and from the back-end Grid-facing software. A three-layer service architecture was thus proposed and delivered enabling the clinician to be isolated from detailed aspects of the underlying Grid.

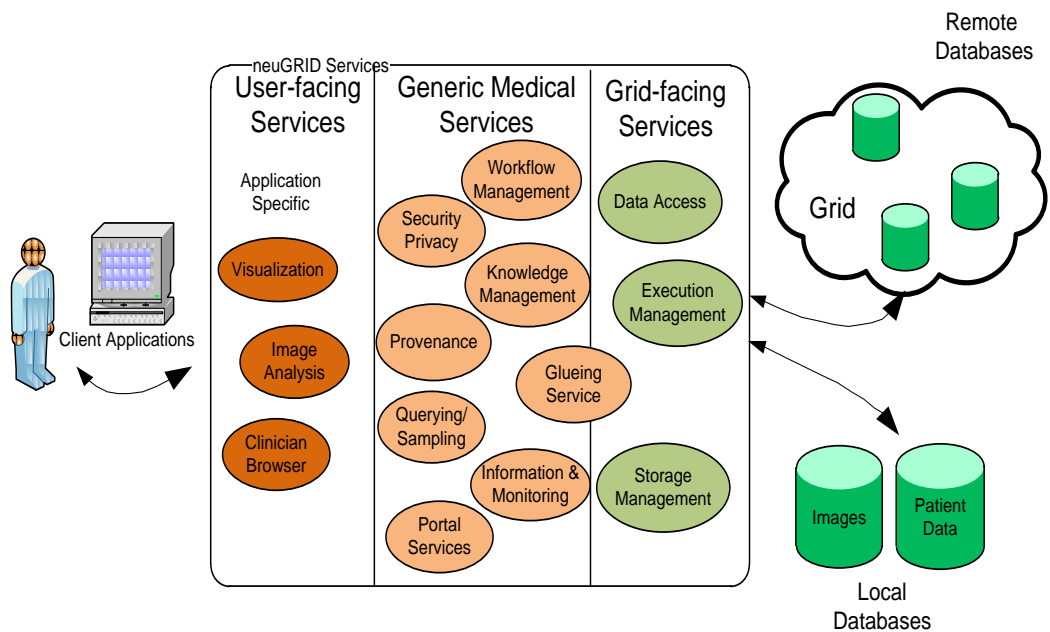

Figure 1: Layered Services Architecture in neuGRID (earlier version in [19]).

The aim of the neuGRID project was to provide an infrastructure and generic services to enable neuroscientists to carry out research required for the study of degenerative brain diseases. At the centre of the neuGRID infrastructure is a distributed computing environment that has been implemented to support the running of complex image processing workflows such as the CIVET algorithm [14], which facilitates the longitudinal measurement of the thinning of the brain cortex. Researchers need to access large distributed libraries of data and to be able to search for a group of images with which they can carry out their analyses based on workflows.

Much clinical research, not only in neuroscience, involves the development of new workflows and image analysis techniques. The ability to construct new workflows or to amend existing ones using established tools is consequently important for clinical research, as outlined in the following section of this paper. neuGRID has followed the MammoGrid design approach in developing a Grid infrastructure based on a SOA [15] to enhance its flexibility and to promote re-usability across other medical domains. This architecture (see Figure 1) allows neuGRID, and now N4U, to be delivered so that its users do not require any advanced Grid knowledge.
By separating Grid specific considerations from those of clinical research, the neuGRID services provide generic and reusable functionality aimed at medical applications. Lower-level services abstract the detail of any specific Grid technology from the upper-level user layers thereby providing application independence and enabling the selection of 'fit-for-purpose' infrastructures. The neuGRID medical services are not specific to a particular application or a particular Grid middleware; they can be used across application domains and can be deployed, potentially, on any Grid (or indeed Cloud) infrastructure.

## 2.2 Data Traceability and the neuGRID Provenance Service

In clinical research environments analyses are often expressed in the form of workflows. A scientific workflow is a step-wise semi-formal specification of a scientific process; one example is that which represents and automates the steps from dataset selection and integration, through analysis, to arrive at final data product presentation and visualization. Workflow management systems support the specification, execution, rerun, and monitoring of scientific processes. Medical and clinical researchers require the ability to track the execution of specified workflows to ensure that important analyses are accurately (and reproducibly) followed. Currently this is carried out manually and can often be error-prone.

A real challenge in the neuroscience scenario is tracking the faults as and when they occur during the workflow specification, distribution and execution phases. This may in turn lead to a loss of user control or repetition of errors during subsequent analyses. Ensuring confidence in the results that are produced is vital if researchers are to accept use of the infrastructure. A number of information points are necessary to allow clinical users and medical researchers to:
  i. Reconstruct a past workflow or parts of it to view the errors at the time of specification.
 ii. Validate a workflow specification against a reference specification.
iii. View the intermediary results produced in the execution of a workflow to determine that those results are valid.
 iv. Validate overall workflow execution results against a reference dataset.
  v. Query information of interest from past analyses.
 vi. Compare different analyses.
vii. Search annotations associated with a pipeline or its components for future reference.

The benefit of managing specified workflows over time is that they can be refined and evolved by researchers collaboratively and can ultimately reach a level of maturity and dependability (the so-called 'gold standard'). To support their analyses, medical researchers need to collect information about (versions of) workflow specifications. These may have been gathered from multiple users and are complimented by whatever results or outcomes were generated. This 'provenance data' is then used as a key driver to improve decision making. Researchers and users must be able to invoke services to monitor and analyse provenance databases to return statistical results that match some criteria as set by the end user. This should provide efficient and dependable problem solving functionality and a reliable decision support system for users.

In neuGRID a Provenance Service has been deployed that captures the information in a provenance database. The service enables the refinement of the workflows in the neuGRID project by capturing (as shown in Figure 2):
a. Workflow specifications.
b. Annotations added to the workflow overall and its constituent workflow tasks.
c. Data or inputs supplied to each workflow task.
d. Links and dependencies between workflow tasks.
e. Output produced by the workflow and each workflow task and
f. Any execution errors generated during analysis.

Figure 2 depicts the process of workflow specification, execution and generation of provenance data. The provenance capture process starts from the authoring step of the workflow. Clinical researchers can write their workflows and analyses in a number of authoring conventions.

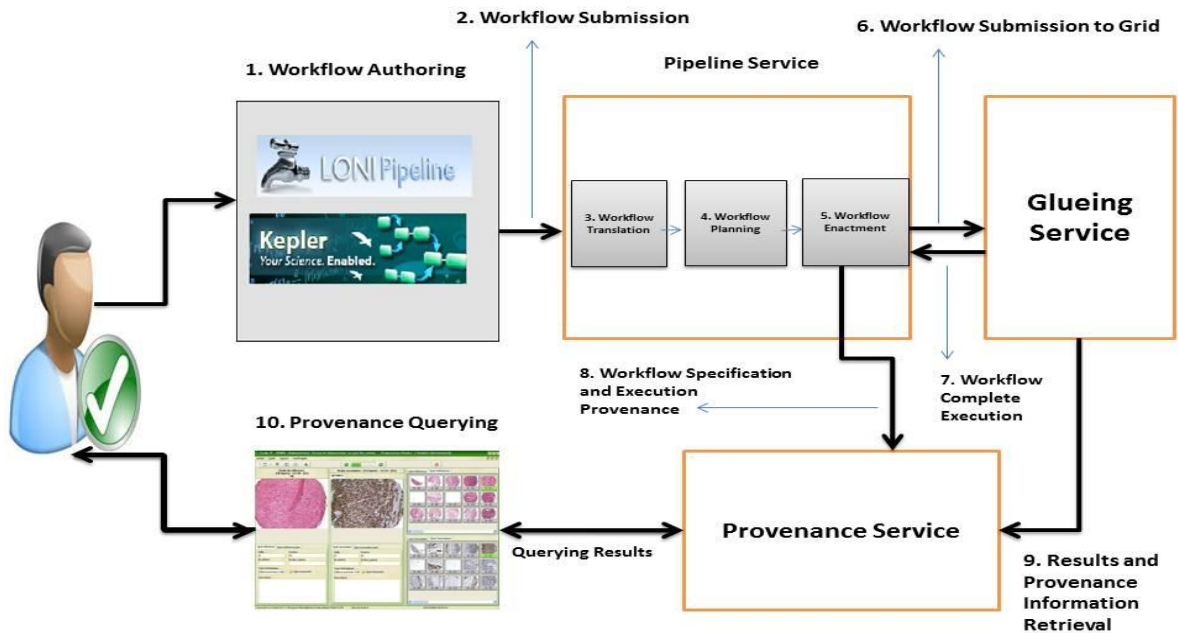

Figure 2: The neuGRID Provenance Service in Neuroimaging Analysis

The Pipeline Service enables the submission of workflows, the tracking of progress and the monitoring of workflows. The user component of the Pipeline Service supports numerous workflow authoring tools (step 1 in Figure 2), such as the LONI Pipeline and Kepler software. When a workflow is submitted to the Pipeline Service (step 2), it is first translated into a common object-oriented workflow format (step 3). The workflow is then forwarded to the Pipeline Service planner for optimisation (step 4) to enable efficient enactment (step 5) and then submitted for execution to the Grid (step 6). The Provenance Service captures the workflow specification as well as any generated execution provenance data as shown in steps 7 and 8 in Figure 2. The results and execution traces retrieved from the Grid (step 9) are also stored by the Provenance Service to have a consolidated view of the full set of specification, execution, monitoring and resultant provenance data. A clinical researcher can then use the querying interface (step 10) of the Provenance Service to browse the provenance information and may even use this information for workflow validation to determine the accuracy of the results.

In practice the Provenance Service is a reusable and extensible service that traces, stores and provides access to analysis data to facilitate improved clinical decisions making. It provides essential support for clinical users to: (a) query analysis information; (b) regenerate analysis workflows; (c) improve and fine-tune workflows; (d) detect errors and unusual behaviour in past analyses; and (e) validate analyses. This means that clinical researchers can have a full and unambiguous logged history of what actions they carried out on which data sets, for what purpose and with which outcome and they can annotate their research processes. This is all made available via an analysis service thereby ensuring the reproducibility and accurate repeatability of their research and they can share those analyses with other medical researchers. As a consequence the provenance database becomes a rich knowledge base of accumulated analysis information and outcomes for clinical and medical researchers to consult as and when required. The next section outlines how software called CRISTAL has been adapted to provide neuGRID provenance tracking functionality for medical analysis.

## 3. Results and Technological Outcomes
### 3.1 CRISTAL as the Basis of a Provenance Service in neuGRID and N4U

The neuGRID Provenance Service is built on a system called CRISTAL [16] that was developed by the authors to manage the construction of large-scale physics detectors for the Large Hadron Collider (LHC) at CERN. CRISTAL can best be described as a distributed data and workflow management system that uses an extendable storage repository and a multi-layered architecture for its component abstraction and object modelling for the design of its components. These techniques are central to handling the complexity of data and workflow traceability in distributed systems and to provide the level of flexibility required to handle the evolving scenarios typical of any research system.

CRISTAL is based on a description-driven approach in which all logic and data structures are 'described' by meta-data, which can be modified and versioned online as the description of the object, component, item or an application changes. The so-called Description-Driven System (DDS) architecture, as advocated previously in [17] is an example of a reflective meta-layer architecture (as shown in Figure 3). CRISTAL makes use of meta-objects to store domain-specific system descriptions (such as items, processes, lifecycles, goals and outcomes), which facilitate orchestration and management of the lifecycles of essential domain objects. In neuroimaging data analysis these objects might be, for example, raw image datasets, pipelines, derived datasets or analysis outcomes etc. As objects, reified system descriptions of DDSs can be organised into libraries that conform with frameworks for modelling of languages, and to their adaptation for specific application domains. An early version of the CRISTAL software (V1.0) has been commercialised and is being applied to, amongst other domains, business process management in the retail, government and logistics domains. The latest version, on which this research is based (V3.0), is in the process of being made available as Open Source.

The meta-data along with the instantiated elements of data are stored in the CRISTAL database and the evolution of the design is tracked by versioning any changes in the meta-data over time. The CRISTAL DDS makes use of meta-objects to store domain-specific system descriptions that manage the lifecycles of application domain objects. The separation of descriptions from their instances enables specification and management to evolve independently and asynchronously. This separation is essential to handle complexity issues that face many web-based computing applications and it facilitates interoperability, reusability and system evolution. Separating descriptions from their instantiation allows new versions of defined objects (and in turn their descriptions) to coexist with older versions. Neuroimaging researchers are constantly developing new algorithms and workflows which may require variations to the provenance data that is collected. At the same time to be useful provenance needs to remain consistent over time, to be traceable, queryable and easily accessible and scientists' analyses need to be conducted on those data. CRISTAL handles all of this. As the detail of the internals of CRISTAL is beyond the brief of this paper the reader is directed to previous publications on DDS (i.e. [17] and [18]) for further detail on DDS theory and the internals of CRISTAL.

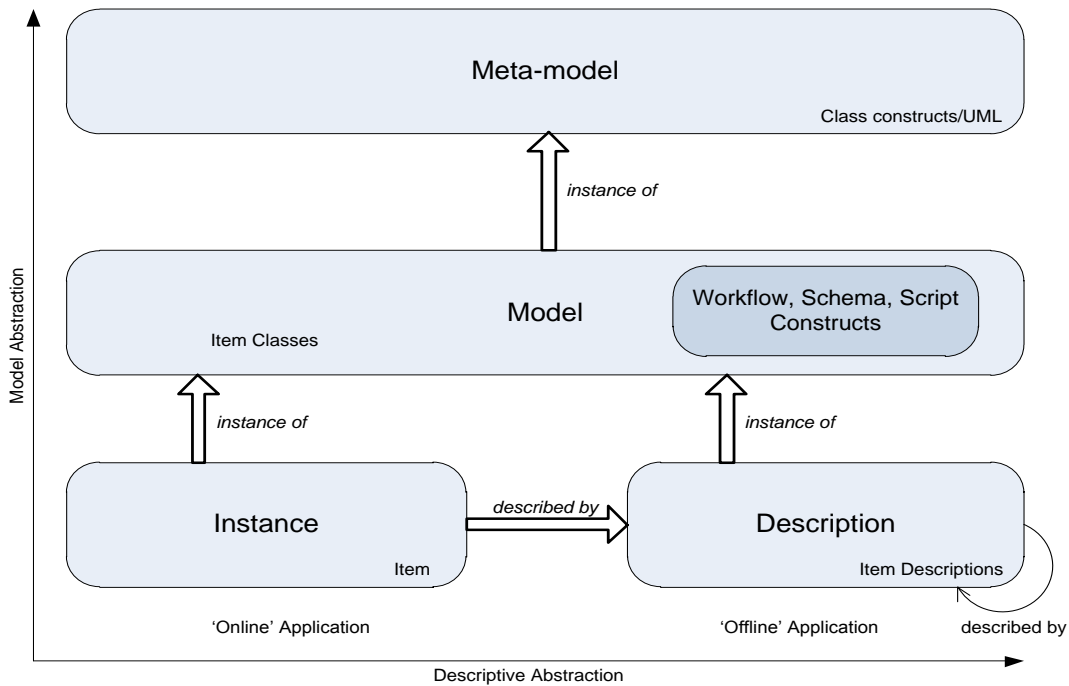

Figure 3: Models and Description in CRISTAL

Using facilities for description capture and dynamic modification, CRISTAL can support modifiable and reconfigurable workflows, such as are present in medical research processes. CRISTAL uses the description-driven nature of its models [17] to enable it to act dynamically on process instances already running and it can intervene in the actual process instances during execution. The workflow activity is actually represented internally in the CRISTAL model as a directed acyclic graph; all associated dependencies, parameters, and environment details being represented in this graph. The schema also provides support to track the workflow evolution with the descriptions of derived workflows and its constituent parts being related to the original workflow activity (consult [19] for details of the implementation of CRISTAL in neuGRID).

Once a workflow has been enacted in the Grid, the Provenance Service coordinates the retrieval of all data outcomes in addition to any data that was produced during the execution of the saved workflow. CRISTAL populates the appropriate structures in the Analysis Base of the N4U project (see [20]) to enable tracking. (The N4U Analysis Base captures information about all image datasets, algorithms, clinical data and accumulated provenance data for a community of researchers). In the CRISTAL model, a workflow consists of a number of activities, these activities being associated to multiple events and each event is associated with an outcome. An activity event may denote that the activity has failed and its outcome may include associated error logs, while the outcome of a successful activity may be the data produced during the runtime of the job. The adopted model enables the pervasive tracking of the entire life-cycle of a neuroimaging workflow, from the pre-planned workflow to the final data outcomes.

## 3.2 The neuGRID Provenance Service in Practice

A prototype Provenance Service was deployed during the final stages of neuGRID and evaluated against the requirements from medical users in the neuroimaging community. To thoroughly evaluate it a two-pronged strategy was pursued. This involved functional testing via large-scale data challenges and a more user-focused testing scheme which was based on a clinical researcher defined case study. The results of such evaluations are considered in detail in the next sections and conclusions are drawn regarding lessons learned and directions for future work in this area. Since the completion of neuGRID the Provenance Service has been further refined in the N4U project to ensure that it satisfies the tracking as detailed in the neuroscience research community's requirements [21] and for potential use outside neuroscience.

The Provenance Service consisted of two layers; an API/Translator layer and the CRISTAL layer. The API layer implements the Web Service that serves as entry point to the Provenance Service. It was implemented using Apache Axis 1.4 (see http://ws.apache.org/axis/) and Tomcat 6.2.20 (see http://tomcat.apache.org/). This layer also consists of a so-called Translator component. Figure 4 shows the portal of the neuGRID Provenance Service. The API allows clients to store workflow templates, to create instances of workflows and to update the status of workflow instances; in N4U this is all managed in the Analysis Base alongside the meta-data describing data-sets (further detail can be found in [20]). The Translator component is responsible for converting the workflow passed to the Provenance Service in a standard format into CRISTAL's internal format. It employs a two-pass mechanism for translation. In the first pass, the workflow is queried for information about each activity such as its *TaskName, Executable, Priority* etc. In the second pass, the CRISTAL workflow is constructed using information that has been mined during the first pass.

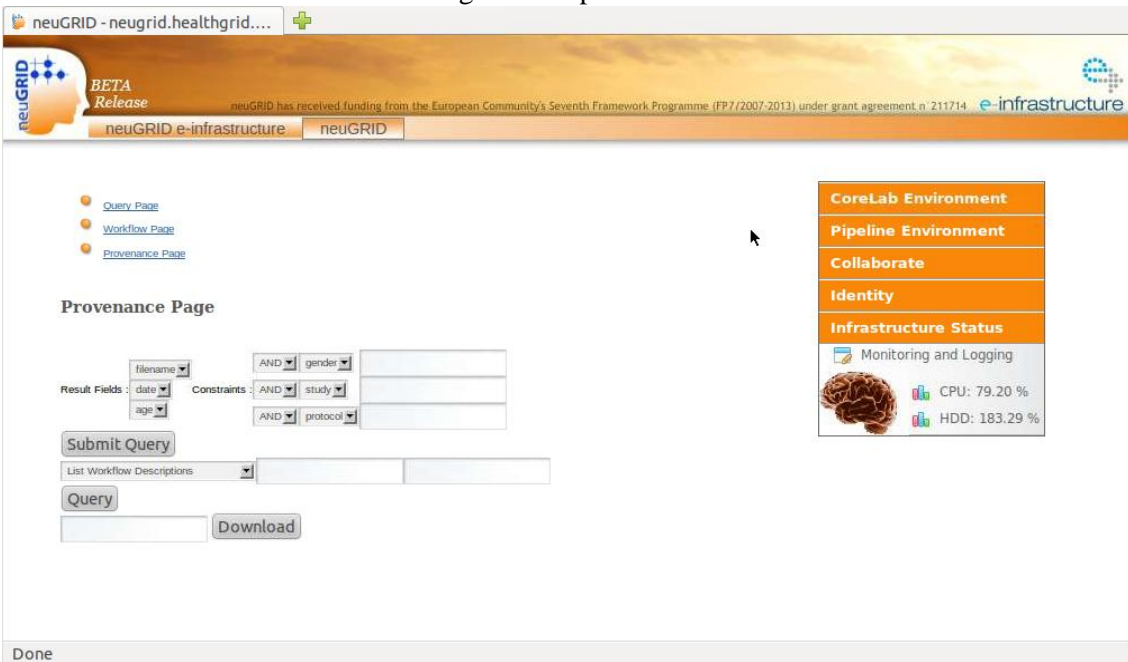

Figure 4: The Provenance Service Portal in neuGRID and N4U.

The CRISTAL layer is the tracking element in the Provenance Service. Once the execution of a workflow commences in the neuGRID infrastructure, a simulation of the workflow is created in parallel in CRISTAL. This enables clients of CRISTAL (principally the Pipeline Service) to send incremental updates to the Provenance Service. The parallel workflow in CRISTAL simulates the execution of the workflow on the Grid infrastructure. The process of adapting CRISTAL for the neuGRID Provenance Service involved creating the appropriate item descriptions within CRISTAL. CRISTAL uses two databases; the 'CRISTAL DB' stores only the CRISTAL internal model to facilitate orchestration of the capture of provenance and the Analysis Base stores the workflows' resulting provenance data, thereby separating the operation of provenance data from its storage and governance. Other neuGRID services (e.g. the Querying Service) directly interact with the Analysis Base for the execution of users' queries to perform search for different datasets, algorithms and pipelines [20].

## 4. Discussion and Evaluation of Outcomes
### 4.1 Evaluation of the neuGRID Platform through Large-Scale Data Challenges

To test the functionality of the neuGRID infrastructure thoroughly and to demonstrate the capability of its services a number of international data challenges were devised. These were designed to test each deployed service individually and to evaluate the neuGRID infrastructure as a whole. They were selected to evaluate the major aspects of the platform including:

a. **Usability and Performance:** that the infrastructure performs efficiently and medical researchers can use its services to perform, trace and validate their studies.
b. **Scalability:** that the infrastructure scales up appropriately in relation to the number of studies and the size of the jobs submitted to it.
c. **Fault tolerance:** that the infrastructure and services should be able to detect, report and recover from any errors that occur.
d. **Functionality:** that the required functionality to successfully carry out and trace the Neuroimaging Analysis is available.

To carry out the evaluation in a structured way, a group of three progressively more intensive data challenges were developed. The first challenge tested the functionality of the underlying Grid infrastructure. In this test, each of the individual services provided by gLite, the middleware used in neuGRID, were tested on the use-case of Figure 6 below [21]. This involved setting up a set of automated scripts that ran standard tests of the services against the ADNI dataset and reported any problems. The test validated system functionality in that all the individual services were reported to work according to the limits of their specifications.

| Experiment duration on the Grid | | < 2 weeks |
|---|---|---|
| Analysed data | Patients | 715 |
| | MRI Scans | 6,235 |
| Number of parallel cores | | 184 |
| Output data produced | | 1 TB |

Table 1: Second Data Challenge Statistics

The second data challenge involved a real world analysis that was performed on the US-ADNI dataset [6]. This challenge was designed to test the infrastructure as a whole via a medium-scale test. Table 1 shows some details about the configuration and results. The dataset used consisted of a total of 6,235 brain scans involving 715 patients. Each scan comprised a 10 to 20 MB data file in MINC format. The entire dataset represented about 108 GB of data. The analysis consisted of running the CIVET pipeline [14] on the dataset and retrieving the results. The CIVET pipeline was converted into an automated workflow using the neuGRID Pipeline Service and then enacted on the underlying gLite infrastructure using the Glueing Service (Figure 2). The analysis steps were tracked, modelled and archived using the Provenance Service. To run the test, two out of three sites (Fatebenefratelli (FBF) in Italy and Karolinska Institute (KI) in Sweden) of the neuGRID infrastructure were commissioned. They were set up with 64-bit worker nodes and 64-bit gridified versions of the CIVET pipeline. Between them, the two sites provided 184 processing cores and 5.3 TB of storage capacity. The primary goal of this experiment was to speed up the process required to measure cortical thickness. For this a gridified version of the CIVET Pipeline was used and a snapshot of the produced results is shown in Figure 5. Areas from red to blue denote increasing thinning of the cortex in a patient with Alzheimer's disease (left) and a normal older person (right).

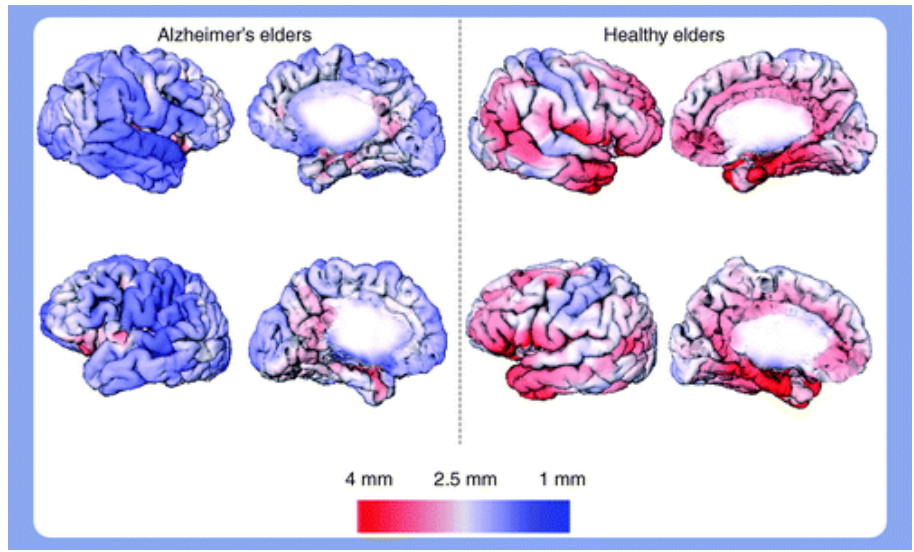

Figure 5: Gridified CIVET pipeline used to assay the cortical thickness marker in neuGRID

Initial tests had shown that CIVET takes about seven hours to process a single scan and as output generates in volume about ten times the input data. The entire test took less than two weeks to complete, as was originally estimated.

| Experiment duration on the Grid | | ~3 months |
|---|---|---|
| Analysed data | Patients | 800 |
|  | MRI Scans | ~7,500 |
| Number of parallel cores | | ~1000 |
| Output data produced | | 2.2 TB |

Table 2: Third Data Challenge

The third data challenge was to test the infrastructure under extreme conditions and to confirm its scalability. Table 2 shows the statistics for the third data challenge. As with the previous challenge, the US-ADNI dataset was chosen. However, compared to 6,235 scans in the second challenge, the third challenge involved analysing approximately 7,500 scans in DICOM format. The scans represented approximately 112 GB of data, with each scan being 10 to 20 MB in size. The challenge consisted of analysing these images using three popular pipelines used by neuroscientists i.e. CIVET [14], FREESURFER [22] and BRAINVISA [23]. A total of 22,500 jobs were to be executed on the infrastructure. This was beyond the capacity of the three neuGRID sites to handle. Therefore, additional sites were added to the Grid for this challenge, made available by the European Grid Infrastructure (EGI). After the neuGRID infrastructure was reconfigured to include these sites, a total of approximately 1000 processing cores became available for this challenge. This reduced the estimated time of completion of the jobs to approximately three months.

The data challenge finished execution in approximately three months and produced a total of 2.2 TB of data. The successful completion of this evaluation confirmed that the infrastructure and services were functioning as anticipated. Moreover, since each workflow instance took a significant amount of time to finish, the collection and preservation of provenance became of paramount importance. It provided valuable information about each step of each workflow instance; the Provenance Service tracked and recorded all this provenance for later retrieval and analysis.

**4.2 In Practice Evaluation of the neuGRID Provenance Service with Neuroscientists**

To evaluate the neuGRID Provenance Service practically with clinical researchers, an end-to-end use-case scenario was developed in collaboration with end users. The community of end users included medically trained clinical researchers, medical statisticians, PhD students and medical imaging experts both from within the neuGRID project and externally. This group participated fully during the design and construction phases of the neuGRID infrastructure.

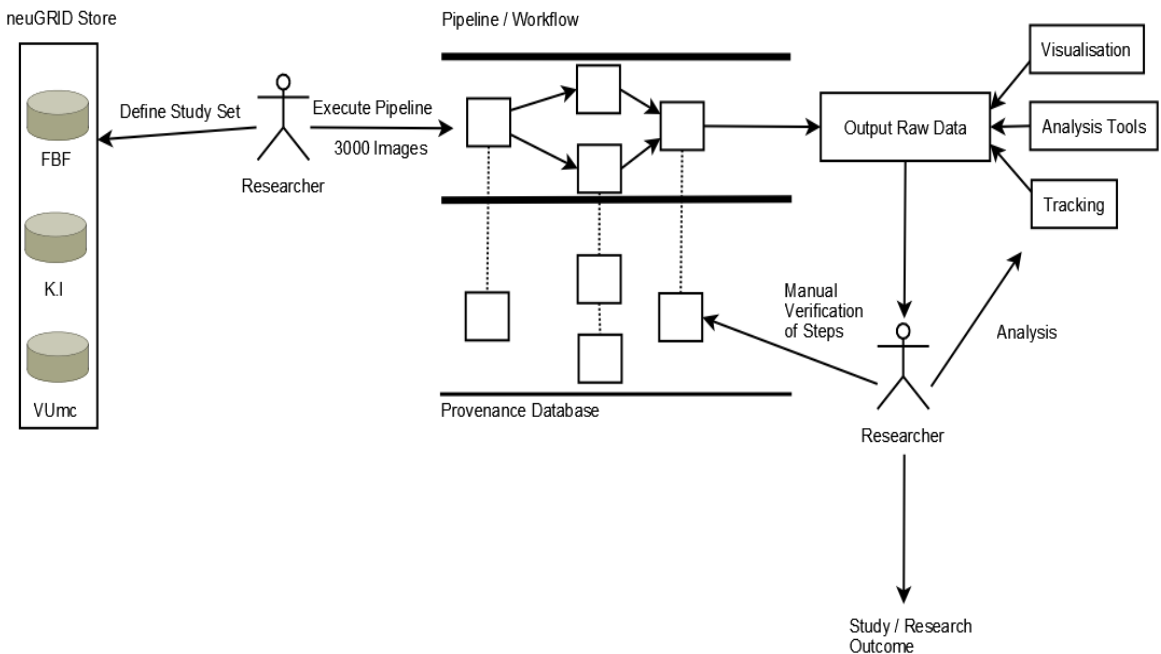

Figure 6: An end-to-end example of the neuGRID Analysis Environment (from [21])

Figure 6 shows the use case which spans a complete analysis cycle in neuGRID from initial data collection, through analysis workflow execution to collaborative data analysis. This use case has been previously introduced briefly in [21] but is expanded here to illustrate the process of evaluation of the Provenance Service. By carrying out test cases that were based on this use case, the neuGRID platform has been thoroughly tested and evaluated by end users. The early stages of the use case initialise the platform prior to analysis being carried out. The first action in the use case, shown on the left in Figure 6 (progression in time is from left to right), was to register the images in the neuGRID data store that have been collected from the candidate hospital acquisition systems or imported from external research projects.

Existing data was subject to quality control, formatting and ethical compliance (plus anonymisation where required). The data was then saved in the neuGRID data model, which enabled other researchers to access it to carry out their research analyses. The registered data was tracked by the Provenance Service, which recorded its creation and stored the definitions of the datasets for subsequent tracking of their usage in analyses.

Once the data had been registered, the next step was to make that data queryable through automated browsing tools. In essence the user interacts with the system using the neuGRID data store, to browse and to identify an appropriately large set of images from a group of hospitals that match the required selection criteria. The Provenance Service is informed of the creation of analysis workflows so that any subsequent usage can be tracked. As an example, a researcher may want to run a comparative analysis using a study set of MRI scans stored in geographically distributed medical centres.

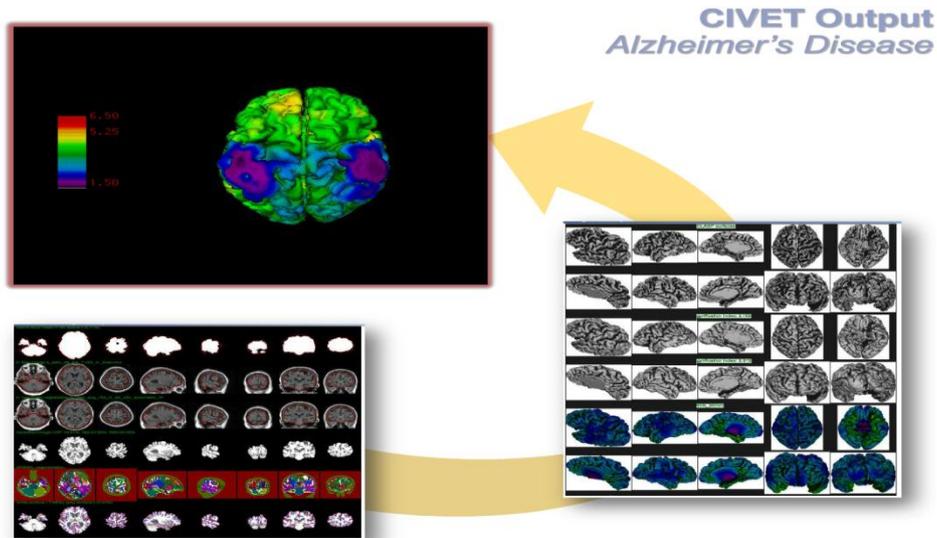

Figure 7: CIVET output in neuGrid

The user would typically interact with the system to choose a study set of perhaps 2000 images (the study is also recorded in the provenance database), may select the CIVET [14] pipeline through which the analysis will take place and would then start the analysis process. An output of such an analysis is shown in Figure 7. Users are not limited to using previously specified workflows and study samples; they can also construct new workflows. These workflows are also captured in the provenance database for later execution and monitoring on the Grid.

It is important that results should be reproduced and reconstructed using their full set of provenance information. It may be required to validate and/or to view the original workflow that had been used to obtain the results. For example, a user may create a new workflow and run it on a test data set. At every stage in workflow execution, the Provenance Service can store any intermediary images or data and a full audit trail of the data is kept. After outcomes and results have been gathered from the analysis, the user can inspect the provenance database to determine that each stage of the analysis has been correctly completed. The results of the user's analysis can then be dumped into that user's preferred analysis tool (such as SPSS [24]) and the complete process can be logged for future reference. Without the mechanism that validates the workflows it would be cumbersome to correct the process and to generate accurate results.

In terms of the benefits of the system using neuGRID, a medical researcher is able to access a powerful test-bed for image analysis, to have access to a set of analysis tools for managing her studies, to log and store algorithms and the assessment of reproducibility and to show the validation of accuracy of the results obtained with a new algorithm are of importance to developers, as well as researchers or clinical end users. The users of neuGRID are able to validate algorithms against 'gold standards', as well as test the sensitivity and inner behaviour of algorithms against normally occurring variations in input data using an imaging database of considerable size.

Standard scientific method requires that research results should be repeatable. In the medical imaging domain repeating experiments is a complex and error-prone process. Medical researchers need to know details such as which images were used, their source, which (versions of) workflows or pipelines were applied, which (versions of) specific algorithms, and settings were used to reproduce the analysis. The level of the traceability that the underlying provenance system gives in neuGRID provides the assurance needed for researchers to validate each others' analyses; these principles can be applied across the domain of medical informatics.

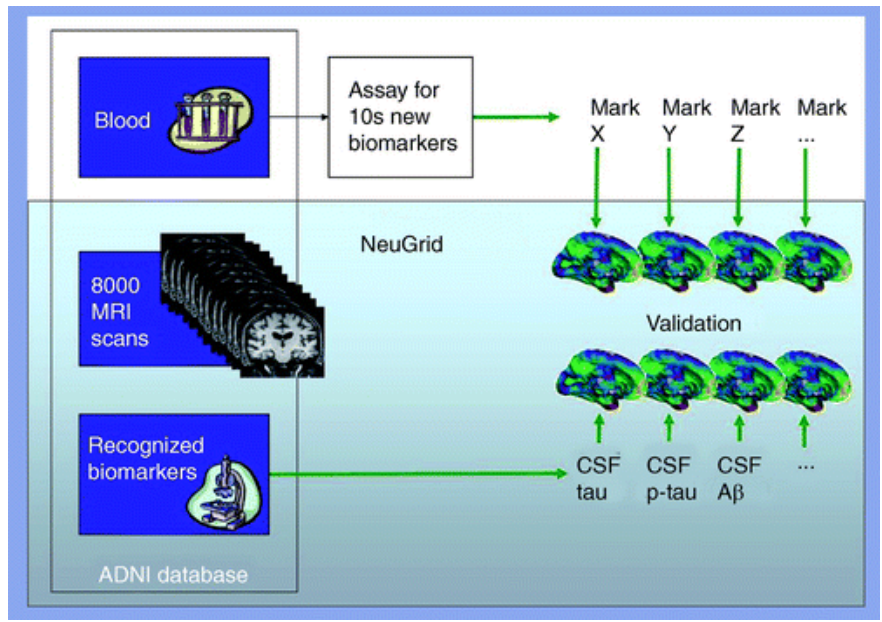

Figure 8: Validation of new biomarkers for Alzheimer's disease using neuGRID

As an example of the use of the neuGRID platform and services consider Figure 8, where a typical research scenario is outlined in which a neuroscientist can rapidly and verifiably evaluate whether a series of newly found assayed bio-markers are improved markers for the onset of Alzheimer's Disease (AD) than existing markers, such as tau, amyloid-β (Aβ) etc. The new markers might be more effective for an early diagnosis of Alzheimer's disease and could also be more accurate in tracking the efficacy of new drug therapies. To carry out her analysis the researcher can use data from the AddNeuroMed and US-ADNI consortia, as stored in the neuGRID Analysis Base and can carry out tests on blood samples from patients with early AD and comparative samples from healthy persons and can save the full provenance of these tests back in the Analysis Base. Through discriminant analysis, the neuroscientist might find that some of the new bio-markers are more accurate than previous markers. At this point, the neuroscientist can access the neuGRID platform to validate the power of the newly discovered markers via imaging indicators of disease progression and *ad-hoc* algorithms. Complex correlative analysis can be carried out on the structural MRI scans taken to study cortical thickness. In doing so, using the Grid capabilities, in a few hours the neuroscientist could demonstrate that a new improved marker has been identified that is actually better than any other previously recognized marker.

NeuGRID concentrated on the management, orchestration and infrastructure aspects of provenance tracking rather on the user-facing services to facilitate usability. However, in practice some medical researchers found the interface to the neuGRID Provenance Service non-intuitive and with only infrequent access, difficult to use. Consequently in the N4U project we are addressing this by the provision of an Analysis Service in which users can define their personalised analyses through a so-called Virtual Laboratory (VL) interface (see Figure 9). The VL comprises the Analysis, Pipeline and Provenance Service plus a 'Dashboard' to present the underlying system to its users, a set of integrated data resources, a set of services enabling access to the neuGRID infrastructure, and a user 'workbench' to define/configure pipelines and inspect analysis output. The Dashboard is the initial point of entry for any user to the underlying N4U services, the work area is the first point of experimentation that empowers the user to use the N4U services in an infrastructure. This Dashboard service presents the underlying system with a standard user-defined look and feel and which can handle role-specific (novice, regular, advanced user) access to the services. This will enable users to prototype their analyses by refining data selections and pipelines, by trying out simple tests and ultimately larger experiments and to visualize the results of those test and experiments. The Analysis Service provides the functionality that is capable of meeting the requirements of the vast majority of scientists working in the field of imaging of neurodegenerative diseases, white matter diseases, and psychiatric diseases. The VL offers scientists access to a wide range of datasets, algorithm applications, access to computational resources, services, and provenance support (see [20]).

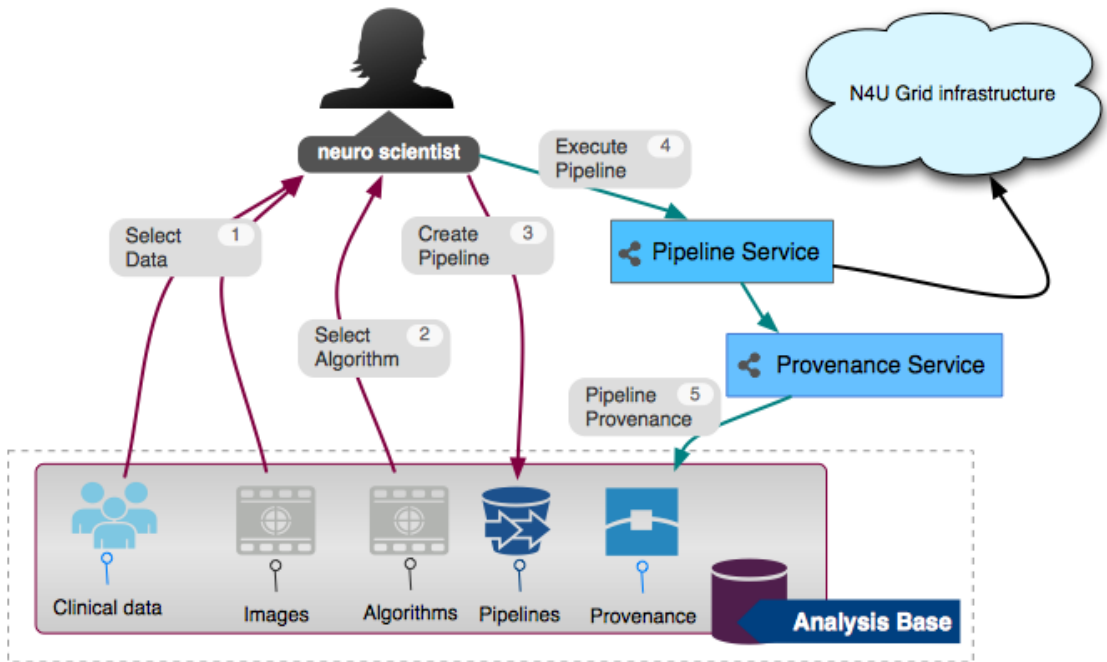

Figure 9: A neuroscientist using the N4U Virtual Laboratory for her analysis

### 4.3 Applicability outside Neuroscience

CRISTAL has been adapted in the Provenance Service to track the provenance of neuroimaging analysis in neuGRID. The important question that immediately comes to mind is how well the Provenance Service can cope with tracking detailed provenance in the medical domain in general. In the neuGRID context, pipelines (workflows) provide a means of capturing processes and their associated metadata, which is a relatively common format to specify and build analyses in many types of medical applications. The underlying CRISTAL structure is flexible, this flexibility enabling CRISTAL to store all of the information that users require in terms of traceability. CRISTAL can therefore capture and link all of the provenance data that is requested by neuroimaging users during the requirements analysis. This was validated in practice through the feedback of neuroimaging users during infrastructure testing when the provenance capture mechanism was demonstrated to them.

All of the processes outlined in the neuroimaging example from neuGRID and N4U are common across the domain of (bio) medical research. The need to capture analysis specifications, both in terms of the data (e.g. images) and the workflows to be run on those data are not specific to neuroimaging. Furthermore the requirement to collect provenance data as workflows are executed and to compile a history or audit of the analysis processes is also common practice especially with the need for repeatability and independent verification. Having worked with mammographers and paediatricians in the research domain it is clear that a general purpose Provenance Service, potentially delivered using CRISTAL, would be widely applicable and desirable to the community of (bio) medical researchers. This hypothesis will be tested in future applications of the neuGRID/N4U Provenance Service.

## 5. Conclusions and Future Directions

This paper has outlined the approach to provenance management that has been developed in the neuGRID and N4U projects. neuGRID has built the foundations for exploitation of Grids in the neuroscience domain through the construction of an adaptable and extensible platform providing customisable and generalised services. N4U is being executed to build the environment in which end-users can access that platform and services and, in particular, to take advantage of the Provenance Service. The major benefit in neuroimaging for Alzheimer's studies is the earlier diagnosis of the disease and the faster development of innovative drugs, that will improve the quality of life of the elderly. neuGRID services have been developed using the service-oriented architecture paradigm as outlined in this paper. These services are reusable both across neuroscience and ultimately for wider application in medical analyses.

In this paper we have demonstrated that the Provenance Service developed for neuGRID can capture the workflow (or pipeline) information that is needed to populate a project-wide database. This service tracks the source of the data and their evolution between different stages of research analyses thereby enabling users to browse analysis information, to edit and generate analysis workflows, to detect and log errors and any unusual behaviour in analyses and, finally, to validate analyses against 'gold-standards'. The use of CRISTAL as the Provenance Service database for neuGRID has enabled neuroscientists to support their complex image analyses over time and to collaborate together in teams with fully versioned workflows and datasets. The work outlined in this paper represents a study carried out with the involvement of clinicians across four medical research centres in Sweden, Italy, Switzerland and the Netherlands whose requirements for analysis provenance have been addressed at a hitherto unprecedented level of granularity. For the first time the medical research community has access to software that enables pan-European collaborative analysis with full data and process traceability and the ability to conduct reproducible cooperative studies on a proven Grid platform that supplies orders of magnitude improvement in the throughput of image analysis.

Ultimately, for the medical informatician, the most important N4U deliverable is the Virtual Laboratory (VL), which provides the environment for users to conduct their analyses on sets of images and associated clinical data derived from a collection of project-specific data. This enables facilities for users to interact with the underlying set of N4U services. All the N4U elements are not specific to neuroscience analysis and can, in principle, be re-used across other domains of medical informatics. In this way we believe that CRISTAL could become an essential building block for future projects requiring data and analysis tracking with provenance management both in execution and design.

**Acknowledgements**
The authors wish to thank their institutes and the European Commission for their support and to acknowledge the contribution of the following neuGRID and N4U project members in the preparation of this paper: clinicians involved were Dr Giovanni Frisoni and Alberto Redolfi (Fatebenefratelli, Brescia), Professor Frederik Barkof and his team at VU Medical Centre (Amsterdam) and Dr Lars-Olaf Wahlund and Eva Orndahl (Karolinska Institute, Stockholm). IT colleagues include David Manset and Jerome Revillard (Maat GKnowledge, France), members of HealthGrid (Clermont, France), and Alex Zijdenbos of Prodema Medical (Braunshofen, Switzerland).